\RequirePackage{fix-cm}
\documentclass[twocolumn,epjc3,final]{svjour3}  
\smartqed  
 \usepackage{mathrsfs}
\usepackage{amsmath, txfonts}
\usepackage{graphicx,bm,booktabs,bm}
\usepackage{subcaption}
\usepackage{longtable,lscape}
\usepackage{ulem} 
\usepackage{overpic,multirow,color}
\definecolor{cover}{rgb}{0.77,0.87,0.88}
\definecolor{blueone}{rgb}{0.1,0.1,.7}
\definecolor{citec}{rgb}{0.14,0.47,0.09}
\definecolor{two}{rgb}{0.0,0.5,0.}
\definecolor{three}{rgb}{.5,.1,0.15}
\usepackage[bookmarks=true,bookmarksopen=false,plainpages=false,breaklinks=true,
   bookmarksnumbered=true,hypertexnames=false,
   filecolor=blue,urlcolor=three,menucolor=three,
   linkcolor=three,citecolor=blueone, colorlinks,
   anchorcolor=blue,runcolor=pink,frenchlinks=red
   pdfstartview=FitH,pdftitle=title,%
   pdfauthor=author]{hyperref}
              \allowdisplaybreaks[4]

\graphicspath{{Figures/}} %

\journalname{Eur. Phys. J. C}
\begin{document}
\title{Roles of $\bar{D}^{*}K^{*}$ and $D^*\bar{D}$ molecular states in decay $B^+ \to D^{*+} D^- K^+$}
\author{Zuo-Ming Ding, Qi Huang \and Jun He\thanksref{e1}
}                     
\thankstext{e1}{Corresponding author: junhe@njnu.edu.cn}
\institute{Department of  Physics and Institute of Theoretical Physics, Nanjing Normal University,
Nanjing 210097, China}

\date{Received: date / Revised version: date}
%
\maketitle

\abstract{

This study investigates the three-body decay process $B^+ \to D^{*+} D^- K^+$,
aiming to explore the possible origins of $T^*_{\bar{c}\bar{s}0}(2870)^0$ and
$\chi_{c1}(3872)$ as intermediate states. Within the molecular
state framework, $T^*_{\bar{c}\bar{s}0}(2870)^0$ and $\chi_{c1}(3872)$ are
considered as possible $\bar{D}^{*}K^{}$ and $D^*\bar{D}$ molecular states,
respectively.  Using effective Lagrangians, the interaction kernels of the
$\bar{D}^{*}K^{*}$ and $D^*\bar{D}$ systems are constructed within the
one-boson-exchange model.  The corresponding rescattering amplitudes and pole
positions are obtained by solving the quasipotential Bethe-Salpeter equation.
These amplitudes are incorporated into the decay amplitude of the three-body
process, and the $D^-K^+$ and $D^{*+}D^-$ invariant mass spectra are simulated
via Monte Carlo methods.  To better reproduce the experimental data, additional
Breit-Wigner contributions from $T^*_{\bar{c}\bar{s}1}(2900)^0$,
$\chi_{c1}(4010)$, and $h_c(4300)$ are included.  The results show a pronounced
enhancement near 2900 MeV in the $D^-K^+$ invariant mass spectrum, strongly
supporting the interpretation of $T^*_{\bar{c}\bar{s}0}(2870)^0$ as a
$\bar{D}^{*}K^{*}$ molecular state.  While the $\bar{D}^{*}K^{*}$ molecular
state provides a reasonable contribution to the $D^-K^+$ spectrum, the
$D^*\bar{D}$ molecular state yields no significant effect on either the $D^-K^+$
or $D^{*+}D^-$ distributions.  This suggests that the observed $\chi_{c1}(3872)$
structure around 3872 MeV may not be interpreted as a $D^*\bar{D}$ molecular
state. } 

\section{Introduction}\label{sec1}

In 2024, the LHCb Collaboration performed a systematic study of the decay
processes $B^+ \to D^{*+}D^-K^+$ and $B^+ \to D^{*-}D^+K^+$~\cite{LHCb:2024vfz}.  Using an amplitude
fit analysis, the invariant mass distributions of both channels were
successfully described.  In the $B^+ \to D^{*+}D^-K^+$ channel, two resonances,
$T^*_{c\bar{s}0}(2870)^0$ and $T^*_{c\bar{s}1}(2900)^0$, were identified in the
$D^-K^+$ invariant mass spectrum, with the measured masses
and widths listed in the first two rows of Table~\ref{LHCbdata}.  These
open-charmed structures were first reported by the LHCb Collaboration in 2020 in
the $B^+ \to D^+D^-K^+$ decay process~\cite{LHCb:2020bls,LHCb:2020pxc}, where
they were named $X_{0,1}(2900)$; the corresponding measurements are given in the
last two rows of Table~\ref{LHCbdata}.  The close agreement of the measured
masses and widths in the two analyses strongly suggests that they correspond to
the same pair of states observed in different decay channels.  

\renewcommand\tabcolsep{0.4cm}
\renewcommand{\arraystretch}{1.5}
\begin{table}[!tb]
\centering
\caption{Masses and widths for the $T^*_{c\bar{s}0}(2870)^0$, $T^*_{c\bar{s}1}(2900)^0$ and $X_{0,1}(2900)$ resonances determined in Refs.~\cite{LHCb:2024vfz,LHCb:2020bls,LHCb:2020pxc}.
\label{LHCbdata}}
\begin{tabular}{ccc }
\hline
Resonance & Mass (GeV) & Width (MeV) \\
\hline
$T^*_{c\bar{s}0}(2870)^0$& 2.887 $\pm$ 0.008 $\pm$ 0.006 & 92 $\pm$ 16 $\pm$ 16 \\
$T^*_{c\bar{s}1}(2900)^0$ & 2.914 $\pm$ 0.011 $\pm$ 0.015 & 128 $\pm$ 22 $\pm$ 23 \\
$X_0(2900)$ & 2.866$ \pm$ 0.007 $\pm$ 0.002 & 57 $\pm$ 12 $\pm$ 4 \\
$X_1(2900)$ & 2.904 $\pm$ 0.005 $\pm$ 0.001 & 110 $\pm$ 11 $\pm$ 4 \\
\hline
\end{tabular}
\end{table}

From the observed decay channel, it is evident that the quark composition of
$T^*_{\bar{c}\bar{s}0}(2870)^0$ and $T^*_{\bar{c}\bar{s}1}(2900)^0$ is
$[\bar{c}\bar{s}u{d}]$, indicating that it is a fully open-flavor exotic hadronic
state. Theoretical studies suggest that these states may correspond to molecular
states~\cite{Liu:2020nil,Ke:2022ocs,Yu:2023avh,Ikeno:2022jbb,Hu:2020mxp,Yang:2024coj,Chen:2020aos},
compact tetraquark states~\cite{Chen:2020aos,Liu:2022hbk,Wang:2020xyc,Zhang:2020oze,Lian:2023cgs}, a
mixture of both structural models~\cite{Albuquerque:2021svg}, or non-hadronic
resonances arising from dynamic effects~\cite{Molina:2023ghu,Burns:2020epm}.
Currently, the structure of these two states remains unattainable, we look
forward to gaining deeper insights through ongoing theoretical studies and
experimental investigations.

Within the aforementioned theoretical framework, the molecular state model has
drawn considerable interest from researchers. This is because the mass of
$T^*_{\bar{c}\bar{s}0}(2870)^0$ is found to
be very close to the $\bar{D}^{*}K^{*}$ threshold, which has inspired many studies
to explore the possibility that they could be molecular states formed by
$\bar{D}^{*}K^{*}$. In fact, even before the experimental observation of
$T^*_{\bar{c}\bar{s}0}(2870)^0$,
Ref.~\cite{Molina:2010tx} had already proposed the possibility of a
$\bar{D}^{*}K^{*}$ molecular structure with $J^P = 0^+$ within the hidden-gauge
formalism. Following the successful detection of the $T^*_{\bar{c}\bar{s}0}(2870)^0$ states, the idea that
it could be interpreted as a $\bar{D}^{*}K^{*}$ molecular state sparked interest and
in-depth discussions among various research groups. Various theoretical
approaches have been employed to study the $T^*_{\bar{c}\bar{s}0}(2870)^0$
resonance and its associated $\bar{D}^{*}K^{*}$
system~\cite{Chen:2021erj,Wang:2021lwy}.

 In our previous studies~\cite{He:2020btl,Kong:2021ohg}, the interaction within the $\bar{D}^{*}K^{*}$ system was
systematically investigated using the quasipotential Bethe-Salpeter equation
(qBSE) approach, and the one-boson exchange model were constructed based on
heavy quark symmetry and chiral symmetry. Meanwhile, in
Ref.~\cite{Kong:2021ohg}, hidden-gauge Lagrangians were adopted to construct the
potential kernels, thereby enhancing the theoretical consistency of the
framework. Both studies consistently indicated that
$T^*_{\bar{c}\bar{s}0}(2870)^0$ can be interpreted as a $\bar{D}^{*}K^{*}$ molecular
state with quantum numbers $J^P= 0^+$. Furthermore, based on the theoretical
framework established in Ref.~\cite{Kong:2021ohg}, we investigated the $B^+ \to
D^{+}D^-K^+$ decay process with $T^*_{\bar{c}\bar{s}0}(2870)^0$ considered as an
intermediate state, and conducted a detailed fit to the corresponding $D^-K^+$
and $D^+D^-$ invariant mass spectrum and Dalitz plot~\cite{Ding:2024dif}. The
analysis demonstrated that, under the assumption that
$T^*_{\bar{c}\bar{s}0}(2870)^0$ is a $\bar{D}^{*}K^{*}$ molecular state, a
satisfactory agreement between the theoretical results and the experimental data
can be achieved. 

Back to the discussion of the decay process $B^+ \to D^{*+}D^-K^+$, another
notable observation is the presence of a distinct structure appearing at
approximately 3.87~GeV in the $D^{*+}D^-$ invariant mass spectrum. The LHCb
collaboration indicates that a $1^{++}$ contribution is required to accurately
describe the distribution of this spectrum. They has tested various effective
models (EFF$_{1++}$), including the exponential non-resonant (NR) amplitude and
a Breit-Wigner-like parametrization for the $\chi_{c1}(3872)$ resonance. The
fitting results indicate that both approaches are capable of describing the
observed structure near 3.87~GeV. However, the LHCb collaboration emphasizes
that the $\chi_{c1}(3872)$ resonance included in the aforementioned fit should
be interpreted only as an effective mode, rather than as a representation of the
genuine $\chi_{c1}(3872) \to D^{*\pm}D^{\mp}$ decay process. Even so, despite
the explanations provided by the LHCb collaboration, considering that the mass
of the $\chi_{c1}(3872)$ resonance is close to 3.87~GeV and lies near the
$D^*\bar{D}$ threshold, we still tend to interpret the observed experimental
structure in this region as a  $D^*\bar{D}$ molecular state, in accordance with many
theoretical
studies~\cite{Liu:2009qhy,AlFiky:2005jd,Wang:2013daa,Dong:2009yp,He:2014nya,Ding:2020dio},
for the purpose of calculating the invariant mass spectra and further evaluating
the feasibility of such assumption.

In this study, we investigate the combined contributions of the $\bar{D}^{*}K^{*}$
and $D^*\bar{D}$ molecular states, corresponding to $T^*_{\bar{c}\bar{s}0}(2870)^0$
and $\chi_{c1}(3872)$, to the $B^+ \to D^{*+}D^-K^+$ decay process.  Other
contributions are modeled as Breit-Wigner resonances.  The $D^-K^+$ and
$D^{*+}D^-$ invariant mass spectra are calculated, and a comparison between our
theoretical predictions and the experimental data from LHCb is performed.  This
analysis aims to improve our understanding of the underlying decay mechanism and
to further clarify the physical nature of the $T^*_{\bar{c}\bar{s}0}(2870)^0$ and
$\chi_{c1}(3872)$ structures.

The remainder of this paper is organized as follows.  Section~\ref{sec2}
presents the formalism for the $B^+ \to D^{*+} D^- K^+$ decay amplitude,
including hadronization, the rescattering mechanism, and the construction of
potential kernels within the qBSE framework.
Numerical results and comparisons with experimental data are discussed in
Section~\ref{sec3}.  Finally, conclusions and implications are summarized in
Section~\ref{sec4}.

\section{Mechanism of  $B^+$$\to$$D^{*+} D^- K^+$ process}\label{sec2}

In this work, we investigate the three-body decay process $B^+ \to D^{*+} D^-
K^+$.  In general, a three-body decay can proceed either through a direct decay
or via intermediate states.  The direct decay mainly represents non-resonant
background contributions.  In the present analysis, these are treated as part of
the background and are not considered explicitly.  Our focus is instead on
processes where the intermediate states arise from rescattering.  Here, $T^*_{\bar{c}\bar{s}0}(2870)^0$ and
$\chi_{c1}(3872)$ are treated as candidate $\bar{D}^{*}K^{*}$ and $D^*\bar{D}$
molecular states, respectively. Other possible intermediate states can be described
phenomenologically using Breit-Wigner parameterizations, as will be discussed
the next section.

In the following, we focus on the explicit rescattering mechanisms considered in our
analysis. The $B^+$ decay is described by a hadronization mechanism  at the quark level.  Next,
within the one-boson-exchange model, we construct the potential kernels and
calculate the rescattering amplitudes for the $\bar{D}^{*}K^{*}$ and $D^*\bar{D}$
systems using the quasipotential Bethe-Salpeter equation framework.  Finally,
these amplitudes are combined with the direct decay amplitude to obtain the
complete expression for the $B^+ \to D^{*+} D^- K^+$ process.

\subsection{Rescattering mechanism}

\begin{figure*}[htbp]
\centering
\subcaptionbox{\label{Fig: diagram2870}}{%
\includegraphics[width=0.45\linewidth, trim=2cm 2cm 2cm 1.5cm, clip]{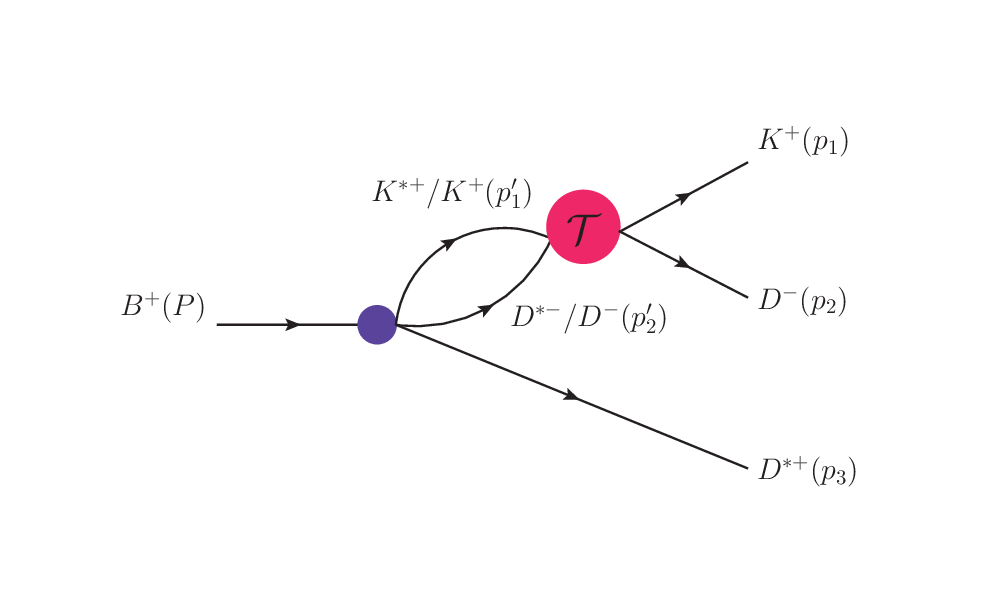}}
\hspace{0.05\textwidth}
\subcaptionbox{\label{Fig: diagram3872}}{%
\includegraphics[width=0.45\linewidth, trim=2cm 1.5cm 2cm 2.5cm, clip]{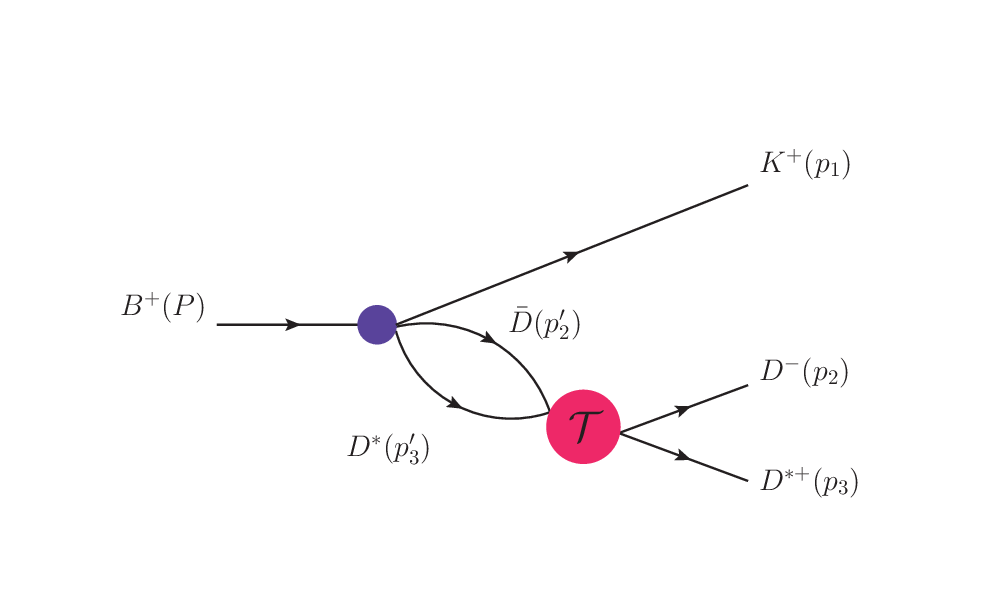}}
\caption{Diagrams for the $B^+ \to D^{*+} D^- K^+$ process involving 
rescattering through (a) $\bar{D}^{*}K^{*}$ and (b) $D^*\bar{D}$ intermediate states.}
\label{Fig: diagram}
\end{figure*}

Within our framework, $T^*_{\bar{c}\bar{s}0}(2870)^0$ and $\chi_{c1}(3872)$ are
characterized as $\bar{D}^{*}K^{*}$ with $J^P=0^+$ and $D^*\bar{D}$ with $J^P=1^+$
molecular states, as illustrated
in Fig.~\ref{Fig: diagram2870} and ~\ref{Fig: diagram3872}, respectively.  The $B^+$ meson first decays into $K^{(*)}$, $\bar{D}^{(*)}$,
 and $D^{*}$, as indicated by the blue circles in Fig.~\ref{Fig:
diagram}. The detailed mechanism of this step will be discussed in the following subsection. Subsequently, the intermediate $\bar{D}^{*}K^{*}$ and $D\bar{D}^*$
systems undergo rescattering, producing the final-state particles $D^-$, $K^+$,
and $D^{*+}$, as marked by the red circles.  In evaluating the rescattering
amplitude $\mathcal{T}$, we include both the $\bar{D}^{*}K^{*}$  
coupling to $D^{-}K^{+}$ as shown in Fig.~\ref{Fig: diagram2870}, as well as the
$D^*\bar{D}$ interaction coupling to $D^{-}D^{*+}$ as shown in Fig.~\ref{Fig: diagram3872}.

\subsection{Hadronization}\label{secHadronization}

\begin{figure}[h!]
  \centering
\includegraphics[width=1.1\linewidth, trim=1.cm 2cm 1.cm 0.5cm, clip]{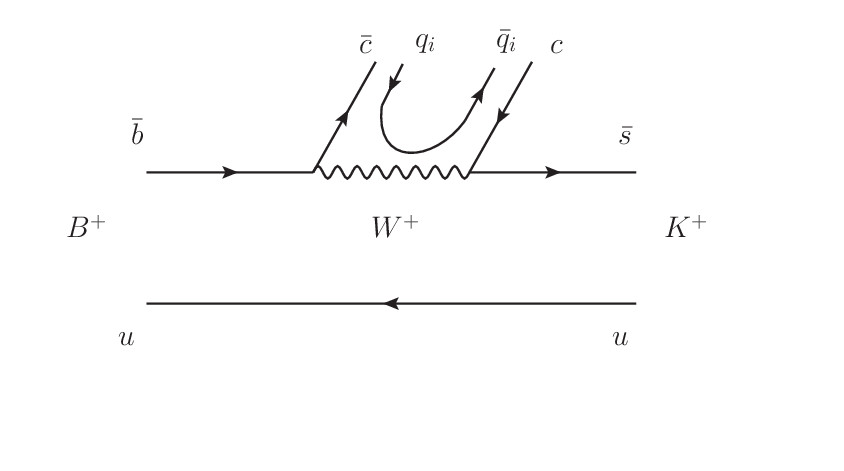}
\caption{Internal emission mechanism at the quark level for $B^+$ decays considered in the current work. \label{Fig: Hadronization}}
\end{figure}

To obtain the amplitude for the decay $B^+ \to D^{*+} D^- K^+$ with rescattering
effects, we first need to specify the decay vertex of the $B^+$ meson. In the
present work, we follow the hadronization procedure outlined in
Ref.~\cite{Oset:2016lyh}.  For the cases considered here, the decay products of
the $B^+$ must contain a $u$ and a $\bar{s}$ quark, a pair
of charm quarks ($c\bar{c}$), and a light quark pair ($q_i\bar{q}_i$).  At the quark
level, there exists only one possible weak decay mechanism for the $B^+$ meson,
as illustrated in Fig.~\ref{Fig: Hadronization}. In this mechanism, the $u$ quark
acts as a spectator, while the $\bar{b}$ quark transforms into a $\bar{c}$ quark
by emitting a $W^+$ boson. The $W^+$ subsequently materializes into a $c$ and a
$\bar{s}$ quark, and an additional light quark pair is generated from the
vacuum.

We note that
this hadronization process only involves internal emission, where an extra
$q_i\bar{q}_i$ pair of the vacuum is added. The subsequent step involves
combining the $q_i\bar{q}_i$ and $c\bar{c}$ pairs to construct meson pairs,
following established hadronization procedures. For this purpose we define the
pseudoscalar ($\mathcal{P}$) and vector ($\mathcal{V}$) matrices:

\begin{equation}
{{\mathcal{P}}} =
\left(
\begin{array}{cccc}
 \frac{\sqrt{3}\pi^0+\sqrt{2}\eta+\eta'}{\sqrt{6}} & \pi^+ & K^+ & \bar{D}^0\\
\pi^- &  \frac{-\sqrt{3}\pi^0+\sqrt{2}\eta+\eta'}{\sqrt{6}}  & K^0 & D^- \\
K^- & \bar{K}^0 & \frac{-\eta+\sqrt{2}\eta'}{\sqrt{3}} & D_s^- \\
D^0 & D^+ & D_s^+ & \eta_c\\
\end{array}
\right),\, \label{PmatrixDK}
\end{equation}
and
\begin{equation}
{\mathcal{V}} =
\left(
\begin{array}{cccc}
 \frac{\rho^0+\omega}{\sqrt{2}} & \rho^+ & K^{* +} & \bar{D}^{* 0} \\
\rho^- & \frac{-\rho^0+\omega}{\sqrt{2}}
 & K^{* 0} & {D}^{* -} \\
K^{* -} & \bar{K}^{* 0} & \phi & D_s^{* -} \\
D^{* 0} & D^{* +} & D_s^{* +} & J/\psi\\
\end{array}
\right)\, .\label{VmatrixDK}
\end{equation}

During the hadronization of the $ c\bar{c} $ pair, it is necessary to produce a vector meson and a pseudoscalar meson. The sequence of this production is of particular importance, as it leads to either a $\mathcal{P}\mathcal{V}$ or $\mathcal{V}\mathcal{P}$ combinations. Therefore, this distinction must be carefully considered in the subsequent analysis. These two combinations can be explicitly expressed as:
\begin{align}
\mathcal{P}\mathcal{V}: \quad c \bar{c} \rightarrow \sum_i c \bar{q}_i q_i \bar{c}&=\sum_i  \mathcal{P}_{4 i}  \mathcal{V}_{i 4}=( \mathcal{PV})_{44}\nonumber\\&=D^0 \bar{D}^{* 0}+D^{+} D^{*-}+\cdots,\\
\mathcal{V}\mathcal{P}: \quad c \bar{c} \rightarrow \sum_i c \bar{q}_i q_i \bar{c}&=\sum_i  \mathcal{V}_{4 i}  \mathcal{P}_{i 4}=( \mathcal{VP})_{44}\nonumber\\&=D^{* 0} \bar{D}^0+D^{*+} D^{-}+\cdots,
\end{align}
the notation "$\cdots$" denotes additional terms that do not overlap with the $D\bar{D}^*$ components. The combination $\mathcal{V}\mathcal{P}-\mathcal{P}\mathcal{V}$ is
\begin{equation}
\mathcal{V}\mathcal{P}-\mathcal{P}\mathcal{V}=D^{*+} D^{-}+D^{* 0} \bar{D}^0-D^{*-} D^{+}-\bar{D}^{* 0} D^0.\label{Hadronization}
\end{equation}
It is evident that the combination $\mathcal{V}\mathcal{P} - \mathcal{P}\mathcal{V}$ uniquely and precisely corresponds to twice the $D\bar{D}^*$ wave function with quantum numbers $I = 0$ and $C = +$, which can be expressed as~\cite{He:2014nya}:

\begin{align}
\left|X_{D^*\bar{D}}^0\right\rangle_{I=0,C=+}= & 
\frac{1}{2}\left[\left(\left|D^{*+} D^{-}\right\rangle+\left|D^{* 0} \bar{D}^0\right\rangle\right)
\right.\nonumber\\
& \left.-\left(\left|D^{+} D^{*-}\right\rangle+\left|D^0 \bar{D}^{* 0}\right\rangle\right)\right]. \label{DD^*}
\end{align}
Accordingly, in the following analysis, only these specific quantum numbers will be taken into account in the rescattering process.

We now turn to the case where the $\bar{D}^{*}K^{*}$ system associated with the
$T^*_{\bar{c}\bar{s}0}(2870)^0$ acts as an intermediate state.  The internal
emission mechanism follows the same pattern as shown in Fig.~\ref{Fig:
Hadronization}, with the only modification being the replacement of the $K^+$
meson by a $K^{*+}$ meson.  To account for the final $K^+D^-$ channel, we
consider the coupled $\bar{D}^*K^*-\bar{D}K$ system.  The isospin wave functions
of the $\bar{D}^{(*)}K^{(*)}$ states can be written
as,~\cite{Huang:2020ptc,Dai:2022htx} 
\begin{align} & \left|\bar{D}^{(*)}K^{(*)}
, I=0\right\rangle=\frac{1}{\sqrt{2}}\left[D^{(*)-}K^{(*)+} -\bar{D}^{(*)
0}K^{(*) 0} \right], \nonumber\\ & \left| \bar{D}^{(*)}K^{(*)},
I=1\right\rangle=\frac{1}{\sqrt{2}}\left[D^{(*)-}K^{(*)+}+\bar{D}^{(*) 0}K^{(*)
0} \right].\label{K^*D^*} \end{align} 
Unlike the $D\bar{D}^*$ system, whose quantum numbers are fixed by hadronization
to $I=0$ and $C=+$, the $\bar{D}^{(*)}K^{(*)}$ system is not subject to such
restrictions.  Thus, the $T^*_{\bar{c}\bar{s}0}(2870)^0$ could correspond to a
$\bar{D}^{*}K^{*}$ molecular state with either $I=0$ or $I=1$.  In the present
study, however, we consider only the final $K^+D^-$ configuration.  Since no
$K^{(*)0}$ can be produced in the $B^+$ decay under consideration, only the
$D^{(*)-}K^{(*)+}$ channels are relevant for evaluating the amplitudes of $B^+ \to
D^{*+} D^- K^+$.  From Eq.~(\ref{K^*D^*}), if the amplitudes for the $I=0$ and
$I=1$ components are denoted by $t_0$ and $t_1$, respectively, the total
contribution from the $D^{-}K^{+}$ channel is $(t_0+t_1)/\sqrt{2}$.

In the above discussion, we have addressed the flavor structure of the decay
vertex.  To construct the full decay amplitudes, we must also specify the
Lorentz structure of the vertex $B^+ \to D^{+} D^{(*)-} K^{(*)+}$, represented by
the solid blue circle in Fig.~\ref{Fig: diagram2870}.  Following the formalism in
Ref.~\cite{Braaten:2004fk}, the Lorentz structure of the amplitude is determined
by Lorentz invariance.  For the specific decay channel $B^+ \to D^{*+} D^- K^+$,
the amplitude can be written as 
$A^\mu_{B^+ \to D^{*+} D^- K^+} = c_1\epsilon^{(1)\mu}$
 where $\epsilon^{(1)\mu}$ is the polarization vector of the
$D^{*+}$ meson.
For the decay $B^+ \to D^{*+} D^{*-} K^{*+}$, the vertex takes the form $A^\mu_{B^+
\to D^{*+} D^{*-} K^{*+}} = c_2\varepsilon^{\mu\nu\alpha\beta} ,
\epsilon_\nu^{(1)} \epsilon_\alpha^{(2)} \epsilon_\beta^{(3)}$, where
$\epsilon^{(1)}$, $\epsilon^{(2)}$, and $\epsilon^{(3)}$ denote the polarization
vectors of the $K^{*+}$, $D^{*-}$, and $D^{*+}$, respectively.  The parameters
$c_1$ and $c_2$ are the corresponding coupling strengths.  Due to the absence of
experimental constraints, both $c_1$ and $c_2$ are treated as free parameters in
our analysis.
Similarly, for the direct decay $B^+ \to D\bar{D}^* K^+$, represented by the
solid blue circle in Fig.~\ref{Fig: diagram3872}, the vertex is ${\cal
A}^\mu_{B^+ \to D^*\bar{D} K^+} = c_3 \epsilon^{(4)\mu}$, where
$\epsilon^{(4)\mu}$ is the polarization vector of the ${D}^*$ within the
$D^*\bar{D}$ molecular state, and $c_3$ is the associated coupling constant.
From Eq.~(\ref{Hadronization}), the hadronization mechanism implies equal
production rates for $D^{*+}$ and $D^-$.  Consequently, the coefficients for the
$D^*\bar{D}$ and $D^{-}K^+$ terms should be identical.  According to Eqs.~(\ref{DD^*})
and (\ref{K^*D^*}), the $D^*\bar{D}$ contribution acquires a factor $1/2$, while the
$D^{-}K^+$ contribution is weighted by $1/\sqrt{2}$.  If the coefficient of
$A_{B^+ \to D^{*+} D^- K^+}$ is $c_1$, then the relation $c_3 = \sqrt{2}c_1$
follows for ${\cal A}_{B^+ \to D^*\bar{D} K^+}$.

\subsection{Potential kernel for rescattering}

The next step involves constructing the potential kernel ${\cal T}$ for the
rescattering process, as the solid red circle shown in Fig.~\ref{Fig:
diagram2870} and Fig.~\ref{Fig: diagram3872}, with the aim of identifying the
pole in the complex energy plane within the qBSE framework and calculating the
corresponding invariant mass spectrum. 

The potential kernel will be calculated within the framework of the
one-boson-exchange model. We begin by considering the $\bar{D}^*K^*$ system, in
which the interaction between the $\bar{D}^{(*)}$ and $K^{(*)}$ mesons is mediated
through the exchange of light mesons $\pi$, $\eta$, $\eta'$, $\rho$, and
$\omega$.  For the systems investigated in this study, the coupling strengths
between the exchanged light mesons and the charmed as well as strange mesons are
required. Consequently, the hidden-gauge Lagrangians incorporating SU(4)
symmetry provide a suitable framework for constructing the interaction
potential, which is given by,~\cite{Bando:1984ej,Bando:1987br,Nagahiro:2008cv}
\begin{align} \label{Eq: lagrangianDK}
 \mathcal{L}_{\mathcal{PP}\mathcal{V}} &=-ig~ \langle V_\mu[\mathcal{P},\partial^\mu \mathcal{P}]\rangle,\nonumber\\
 \mathcal{L}_{\mathcal{VV}\mathcal{P}} &=\frac{G'}{\sqrt{2}}~\epsilon^{\mu\nu\alpha\beta}\langle\partial_\mu \mathcal{V}_\nu \partial_\alpha \mathcal{V}_\beta \mathcal{P}\rangle, \nonumber\\
 \mathcal{L}_{\mathcal{VV}\mathcal{V}}&=ig ~\langle (\mathcal{V}_\mu\partial^\nu \mathcal{V}^\mu-\partial^\nu \mathcal{V}_\mu \mathcal{V}^\mu) \mathcal{V}_\nu\rangle,
\end{align}
where $G'={3g'^2}/{4\pi^2f_{\pi}}$, $g'=-{G_{\mathcal{V}}m_{\rho}}/{\sqrt{2}{f_{\pi}}^2}$, $G_\mathcal{V}\simeq 55$~MeV and $f_\pi=93$~MeV and the coupling constant $g=M_\mathcal{V}/{2f_{\pi}}$, $M_\mathcal{V}\simeq 800$~MeV~\cite{Nagahiro:2008cv}.
The $\mathcal{P}$ and $\mathcal{V}$ are the pseudoscalar and vector matrices under SU(4) symmetry as Matrices.~(\ref{PmatrixDK}) and ~(\ref{VmatrixDK}).

We now proceed to discuss the $D^*\bar{D}$ system. As discussed in
Ref.~\cite{Ding:2024dif}, for hidden heavy states, cross diagrams emerge in the
$D^*\bar{D}$ cases due to the coupling between components containing charm
$c$ quark [the last two components in Eq.~(\ref{DD^*})] and those without [the first two
components in Eq.~(\ref{DD^*})]. The mesons participating in the reaction
typically differ between direct and cross diagrams. For the direct diagram,
contributions from vector mesons ($\rho$ and $\omega$) exchange are considered.
For the cross diagram, contributions from vector mesons ($\rho$ and $\omega$)
and pseudoscalar meson ($\pi$ and $\eta$) exchange are considered. Different
from the $\bar{D}^{(*)}K^{(*)}$ channel, two charmed mesons are involved here.
Hence, heavy quark symmetry is more suitable to describe the interaction. The
heavy quark effective Lagrangian for heavy mesons interacting with light mesons
reads~\cite{Cheng:1992xi,Yan:1992gz,Wise:1992hn,Burdman:1992gh,Casalbuoni:1996pg},

\begin{align}
 \mathcal{L}_{\mathcal{P}^* \mathcal{PP} }= & -\frac{2 g}{f_\pi}\left(\mathcal{P}_b \mathcal{P}_{a \lambda}^{* \dagger}+\mathcal{P}_{b \lambda}^* \mathcal{P}_a^{\dagger}\right) \partial^\lambda \mathbb{P}_{b a} \nonumber\\
& +\frac{2 g}{f_\pi}\left(\widetilde{\mathcal{P}}_{a \lambda}^{* \dagger} \widetilde{\mathcal{P}}_b+\widetilde{\mathcal{P}}_a^{\dagger} \widetilde{\mathcal{P}}_{b \lambda}^*\right) \partial^\lambda \mathbb{P}_{a b}, \\
 \mathcal{L}_{\mathcal{P}^*\mathcal{PV}}
  =&- 2\sqrt{2}\lambda{}g_V v^\lambda\varepsilon_{\lambda\mu\alpha\beta}
  (\mathcal{P}^{}_b\mathcal{P}^{*\mu\dag}_a +
  \mathcal{P}_b^{*\mu}\mathcal{P}^{\dag}_a)
  (\partial^\alpha{}\mathbb{V}^\beta)_{ba}\nonumber\\
&-  2\sqrt{2}\lambda{}g_V
v^\lambda\varepsilon_{\lambda\mu\alpha\beta}
(\widetilde{\mathcal{P}}^{*\mu\dag}_a\widetilde{\mathcal{P}}^{}_b
+
\widetilde{\mathcal{P}}^{\dag}_a\widetilde{\mathcal{P}}_b^{*\mu})
  (\partial^\alpha{}\mathbb{V}^\beta)_{ab},\label{Eq:lagrangianDD}
\end{align} 
where the velocity $v$ should be replaced by
$i\overleftrightarrow{\partial}/2\sqrt{m_i m_f}$ with $m_{i,f}$ being the mass
of the initial or final heavy meson. ${\mathcal{P}}^{(*)T} = (D^{(*)0}, D^{(*)+},
D_s^{(*)+})$. The third $\mathcal{P}$ or $\mathcal{V}$ in the subscript of $\mathcal{L}$ denotes an SU(3) pseudoscalar or vector matrix, whose elements
are the first $3\times3$ elements of the SU(4) pseudoscalar or vector matrix in Matrix.~(\ref{PmatrixDK}) or (\ref{VmatrixDK}), respectively. 
The parameters involved here were determined in the literature as $g=0.59$, $\lambda=0.56$~GeV$^{-1}$, $g_V=5.9$ and  $f_\pi=132$ MeV~\cite{Liu:2009qhy,Falk:1992cx,Isola:2003fh,Chen:2019asm}.

Besides, contribution from the $J/\psi$ exchange is also considered in the
current work since it is found important in the interaction between charmed and anticharmed
mesons~\cite{Ding:2023yuo}. The Lagrangians are written with the help of heavy quark effective
theory as \cite{Casalbuoni:1996pg,Oh:2000qr}, 
\begin{align}
	{\cal L}_{D^*\bar{D}J/\psi}&=
g_{D^*D\psi} \,  \, \epsilon_{\beta \mu \alpha \tau}
\partial^\beta \psi^\mu (\bar{D}
\overleftrightarrow{\partial}^\tau D^{* \alpha}+\bar{D}^{* \alpha}
\overleftrightarrow{\partial}^\tau D) \label{Eq:jpsi}, 
\end{align}
 where the
couplings are related to a single parameter $g_2$ as
${g_{DD\psi}}/{m_D}= 2 g_2 \sqrt{m_\psi }$,with
$g_2={\sqrt{m_\psi}}/({2m_Df_\psi})$ and $f_\psi=405$~MeV.

With the above Lagrangians, the potential kernel  in the one-boson-exchange
model can be constructed by applying the standard Feynman rules, and is
expressed as
\begin{equation}%
{\cal V}_{\mathcal{P(V)}}=I^{d,c}_i\Gamma_{1(\mu)}\Gamma_{2(\nu)}  P^{(\mu\nu)}_{\mathcal{P(V)}}f(q^2),\label{V}
\end{equation}
for pseudoscalar ($\mathcal{P}$) and vector ($\mathcal{V}$) exchange,
respectively.  Here, $\Gamma_1$ and $\Gamma_2$ are respectively used to
represent the upper and lower vertices of the one-boson-exchange Feynman
diagram. The term $I_{i}^{d,c}$ represents the flavor factor associated with a
specific meson exchange, which can be derived using the Lagrangians given in
Eqs.~(\ref{Eq: lagrangianDK}), (\ref{Eq:lagrangianDD}), and (\ref{Eq:jpsi}), as
well as the matrices presented in Matrices.~(\ref{PmatrixDK}) and
(\ref{VmatrixDK})~\cite{He:2015mja}. The superscripts ``$d$" and ``$c$" denote
contributions from the direct and crossed diagrams, respectively. Explicitly,
the flavor factors are $I_{\pi}^d=-3/2$, $I_{\eta}^d=0$, $I_{\eta'}^d=1/2$,
$I_{\rho}^d=-3/2$, and $I_{\omega}^d=1/2$ for the $\bar{D}^*K^*-\bar{D}K$
rescattering process, while for the ${D}{\bar{D}^*}$ rescattering, we have
$I_{\rho}^d=3/2$, $I_{\omega}^d=1/2$, $I_{J/\psi}^d=1$, $I_{\rho}^c=-3/2$,
$I_{\omega}^c=-1/2$, $I_{\pi}^c=-3/2$, $I_{\eta}^c=-1/6$, and $I_{J/\psi}^c=-1$.
The propagators are defined as
\begin{align}
P_{{\mathcal{P}}}= \frac{i}{q^2-m_{{\mathcal{P}}}^2},\ \
P^{\mu\nu}_{\mathcal{V}}=i\frac{-g^{\mu\nu}+q^\mu q^\nu/m^2_{\mathcal{V}}}{q^2-m_{\mathcal{V}}^2},
\end{align}
where $q$ denotes the momentum of the exchanged meson and $m_{\mathcal{V,P}}$
represents its mass. A form factor of the form
$f(q^2)=\Lambda_e^2/(q^2-\Lambda_e^2)$, with a cutoff parameter $\Lambda_e$, is
introduced to account for the off-shell effects of the exchanged meson. This
type of form factor was also employed in Ref.~\cite{Gross:2008ps} to study
nucleon-nucleon scattering within the spectator approximation framework, which
is analogous to the approach used in the present work. It serves to prevent
overestimation of the contribution from $J/\psi$ exchange in $D^*\bar{D}$
rescattering within the current model.

\subsection{Rescattering amplitudes in qBSE approach} \label{secamplitudes}

The amplitude of the rescattering process is calculated by substituting the
previously derived potential kernel into the qBSE
framework~\cite{He:2014nya,He:2015mja,He:2017lhy,He:2015yva,He:2017aps}. In this
and the following subsection, we use the $\bar{D}^{(*)}K^{(*)}$ rescattering
process as an illustrative example, where the two constituents are denoted as
particle~1 and particle~2  as shown in Fig.~\ref{Fig: diagram2870}.  Following the partial wave decomposition, the
four-dimensional qBSE can be formulated as a one-dimensional
form~\cite{Kong:2021ohg,Ding:2024dif,He:2015cea,He:2019ify}. The rescattering amplitude
$\cal T$ corresponding to a specific spin-parity $J^P$ can be expressed
as~\cite{He:2015mja}
\begin{align}
&i{{\cal T}}^{J^P}_{\lambda'_1,\lambda'_2,\lambda_1,\lambda_2}({\rm p}',{\rm p})\nonumber\\
&=i{{\cal V}}^{J^P}_{\lambda'_1,\lambda'_2,\lambda_1,\lambda_2}({\rm p}',{\rm
p})+\frac{1}{2}\sum_{\lambda''_1,\lambda''_2}\int\frac{{\rm
p}''^2d{\rm p}''}{(2\pi)^3}\nonumber\\
&\cdot
i{{\cal V}}^{J^P}_{\lambda'_1,\lambda'_2,\lambda''_1,\lambda''_2}({\rm p}',{\rm p}'')
G_0({\rm p}'')i{{\cal T}}^{J^P}_{\lambda''_1,\lambda''_2,\lambda_1,\lambda_2}({\rm p}'',{\rm
p}),\quad\quad \label{Eq: BS_PWA}
\end{align}
where the indices $\lambda'_1$, $\lambda'_2$, $\lambda''_1$, $\lambda''_2$, $\lambda_1$, and $\lambda_2$ denote the helicities of the two rescattering constituents corresponding to the final, intermediate, and initial particles 1 and 2, respectively. $G_0({\bm p}'')$ represents a reduced propagator expressed in the center-of-mass frame, with $P = (M, {\bm 0})$ defined as
\begin{align}
	G_0&=\frac{\delta^+(p''^{~2}_2-m_2^{2})}{p''^{~2}_1-m_1^{2}}\nonumber\\
  &
  =\frac{\delta^+(p''^{0}_2-E_2({\bm p}''))}{2E_2({\bm p''})[(W-E_2({\bm
p}''))^2-E_1^{2}({\bm p}'')]},
\end{align}
where $m_{1,2}$ denotes the mass of particle 1 or 2. According to the requirements of the spectator approximation, the heavier particle (particle 2 in this case) is placed on-shell, with its four-momentum given by $p''^0_2 = E_{2}({\rm p}'') = \sqrt{m_{2}^{~2} + {\rm p}''^2}$. The corresponding four-momentum component for the lighter particle (particle 1) is then $p''^0_1 = W - E_{2}({\rm p}'')$, where $W$ represents the center-of-mass energy of the system. In this work, unless otherwise specified, we define ${\rm p} = |{\bm p}|$. Furthermore, the momentum vectors are assigned as ${\bm p}''_1 = -{\bm p}''$ for particle 1 and ${\bm p}''_2 = {\bm p}''$ for particle 2.

The partial wave potential ${\cal
V}_{\lambda'_1,\lambda'_2,\lambda_1,\lambda_2}^{J^P}$ can be obtained from the
potential as
\begin{align}
i{\cal V}_{\lambda'_1,\lambda'_2,\lambda_1,\lambda_2}^{J^P}({\rm p}',{\rm p})
&=2\pi\int d\cos\theta
~[d^{J}_{\lambda_{21}\lambda'_{21}}(\theta)
i{\cal V}_{\lambda'_1\lambda'_2,\lambda_1\lambda_2}({\bm p}',{\bm p})\nonumber\\
&+\eta d^{J}_{-\lambda_{21}\lambda'_{21}}(\theta)
i{\cal V}_{\lambda'_1,\lambda'_2,-\lambda_1,-\lambda_2}({\bm p}',{\bm p})],\label{Eq:PWAV}
\end{align}
where $\eta = P P_1 P_2 (-1)^{J - J_1 - J_2}$, with $P$ and $J$ representing the parity and spin of the system and its constituents 1 and 2, respectively. Here, $\lambda_{21} = \lambda_2 - \lambda_1$. The initial and final relative momenta are defined as ${\bm p}' = (0, 0, {\rm p}')$ and ${\bm p} = ({\rm p} \sin\theta, 0, {\rm p} \cos\theta)$, respectively. The term $d^J_{\lambda'\lambda}(\theta)$ denotes the Wigner $d$-matrix. An exponential regularization is incorporated as a form factor into the reduced propagator, expressed as $G_0({\rm p}'') \to G_0({\rm p}'') e^{-2(p''^2 - m_2^2)^2/\Lambda_r^4}$~\cite{He:2015mja}. The cutoff parameters $\Lambda_r$ and $\Lambda_e$ may take different values but yield a similar influence on the results. For simplicity, we set $\Lambda_r = \Lambda_e$ in the present study.

The amplitude ${\cal T}$ can be determined by discretizing the momenta ${\rm p}'$, ${\rm p}$, and ${\rm p}''$ in the integral equation~(\ref{Eq: BS_PWA}) using Gaussian quadrature with a weighting function $w({\rm p}_i)$. Following this discretization procedure, the integral equation is transformed into a corresponding matrix equation~\cite{He:2015mja},
\begin{align}
{T}_{ik}
&={V}_{ik}+\sum_{j=0}^N{ V}_{ij}G_j{T}_{jk},\label{Eq: matrix}
\end{align}
where $i$, $j$, and $k$ denote the indices corresponding to the discretized momenta. The helicities are also incorporated into these indices. The value of $N$ represents the Gaussian discretization dimension, the specific value of which is determined by the stability of the numerical results. In the present study, the discretization dimension is set to 10. The propagator $G$ is expressed in the form of a diagonal matrix:
\begin{align}
	G_{j>0}&=\frac{w({\rm p}''_j){\rm p}''^2_j}{(2\pi)^3}G_0({\rm
	p}''_j), \nonumber\\
G_{j=0}&=-\frac{i{\rm p}''_o}{32\pi^2 W}+\sum_j
\left[\frac{w({\rm p}_j)}{(2\pi)^3}\frac{ {\rm p}''^2_o}
{2W{({\rm p}''^2_j-{\rm p}''^2_o)}}\right],
\end{align}
where the on-shell momentum is given by
\begin{align}{\rm p}''_o=\frac{1}{2W}\sqrt{[W^2-(m_1+m_2)^2][W^2-(m_1-m_2)^2]}.\label{Eq: mometum onshell}
\end{align}
To identify the pole of the rescattering amplitude in the complex energy plane, we search for the location where $|1 - {V}G| = 0$, with $z = E_R + i\Gamma/2$ representing the total energy and decay width of the resonance.

\subsection{Amplitudes with rescattering mechanism}

After incorporating the amplitudes of the direct decay and rescattering
processes, the total amplitude for the reaction $B^+ \to D^{*+} D^- K^+$ with
$\bar{D}^{(*)}K^{(*)}$ rescattering can be formulated in the center-of-mass
frame of particles 1 and 2 as follows~\cite{Ding:2023yuo,He:2017lhy}:
\begin{align}
{\cal M}(p_1,p_2,p_3)&=\sum_{\lambda'_{1},\lambda'_{2}}\int \frac{d^4p'_2}{(2\pi)^4} 
{\cal T}_{\lambda'_{1},\lambda'_{2};\lambda_{1},\lambda_{2}}(p_1,p_2;p'_1,p'_2) \nonumber\\
&\ \cdot \ G_0(p'_2){\cal A}_{\lambda'_{1},\lambda'_{2},\lambda_3}(p'_1,p'_2,p_3).
\end{align}
Here, the momenta have been transformed to the center-of-mass frame of particles 1 and 2.

To analyze the direct decay amplitude ${\cal A}_{\lambda'_1,\lambda'_2,\lambda_3}$, a partial wave expansion is required, analogous to the expansion applied to the rescattering potential kernel ${\cal V}$ presented in Eq.~(\ref{Eq:PWAV}), as detailed in Ref.~\cite{Gross:2008ps}.
\begin{align}
{\cal A}_{\lambda'_{1},\lambda'_{2},\lambda_3}(p'_1,p'_2,p_3)
=\sum_{J,\lambda'_{1},\lambda'_{2}}N_JD^{J*}_{\lambda'_{1},\lambda'_{2}}( \Omega'_2) 
{\cal A}^{J}_{\lambda'_{1},\lambda'_{2},{\lambda}_{3}}({\rm p}'_2,p_3),\label{Eq: decay}
\end{align}
where $N_J$ is a normalization constant with the value of $\sqrt{(2J+1)/4\pi}$. 
$D^{J*}_{\lambda'_{1},\lambda'_{2}}( \Omega'_2) $ is Wigner matrix with $\Omega_2$ being the spherical angle of momentum of particle 2. Hence, the partial-wave amplitude for $J$ is given by
\begin{align}
&{\cal M}^{J}(p_1,p_2,p_3)\nonumber\\
&=N_{J}\sum_{J\lambda'_{1},\lambda'_{2}}D^{J*}_{\lambda_{1},\lambda_{2};\lambda'_{1},\lambda'_{2}}( \Omega_2) 
\int \frac{{\rm p}'^{2}_2d{\rm p}'_2}{(2\pi)^3}\nonumber\\ 
&\  \cdot\  i{\cal T}^{J}_{\lambda_{1},\lambda_{2};\lambda'_{1},\lambda'_{2}}({\rm p}'_2,{\rm p}_2)\cdot G_0({\rm p}'_2) 
{\cal A}^{J}_{\lambda'_{1},\lambda'_{2},\lambda_3}({\rm p}'_2,p_3).
\end{align}

For intermediate states characterized by a definite spin parity $J^P$, the corresponding amplitude ${\cal M}^{J^P}$ can be expressed as 
\begin{align}
{\cal M}^{J^P}(p_1,p_2,p_3)
&=\frac{1}{2}N_{J}\sum_{\lambda'_{21}}
\int \frac{{\rm p}'^{2}_2d{\rm p}'_2}{(2\pi)^3}
i{\cal T}^{J^P}_{\lambda_{1},\lambda_{2};\lambda'_{1},\lambda'_{2}}({\rm p}'_2,{\rm p}_2)
\nonumber\\ 
&\  \cdot\   G_0({\rm p}'_2) 
{\cal A}^{J^P}_{\lambda'_{1},\lambda'_{2},\lambda_3}({\rm p}'_2,p_3),\label{amplitude}
\end{align}
where
\begin{align}
&{\cal A}^{J^P}_{\lambda'_{1},\lambda'_{2},\lambda_3}(\mathrm{p}'_2,p_3)\nonumber\\
&=\int d\Omega'_2 \left[ D^{J*}_{\lambda_{1},\lambda_{2},\lambda'_{1};\lambda'_{2}}(\Omega'_2) 
{\cal A}_{\lambda'_{1},\lambda'_{2},\lambda_3}(\mathrm{p}'_2,p_3)\right.\nonumber\\
&\left.  + \  \eta' D^{J*}_{\lambda_{1},\lambda_{2};-\lambda'_{1},-\lambda'_{2}}(\Omega'_2) 
{\cal A}_{-\lambda'_{1},-\lambda'_{2},\lambda_3}(\mathrm{p}'_2,p_3)\right].
\end{align}

\section{Analysis of invariant mass spectrum of the decay $B^+ \to D^{*+} D^- K^+$}\label{sec3}

\subsection{Invariant mass spectrum} \label{sec3a}

In this section, we present the analysis of the invariant mass spectrum for the
$B^+ \to D^{*+} D^- K^+$ decay process. Based on the formalism described above,
we include the amplitudes from rescattering, ${\cal M}_{\bar{D}^{*}K^{*}}$ and
${\cal M}_{D^*\bar{D}}$, arising from the $\bar{D}^{*}K^{*}$ and $D^*\bar{D}$
molecular states, which are associated with the $T^*_{\bar{c}\bar{s}0}(2870)^0$
and $\chi_{c1}(3872)$, respectively.
In addition, we introduce the amplitudes $\sum{\cal M}_{\rm BW}$ with Breit-Wigner
parameterizations for the $\chi_{c1}(4010)$, $h_c(4300)$, and
$T^*_{\bar{c}\bar{s}1}(2900)^0$ resonances. Background contributions from
different channels are also included, denoted by ${\cal M}_{\rm bk}$.

Hence, the total amplitude can be expressed as
\begin{align}
{\cal M} = {\cal M}_{\bar{D}^{*}K^{*}} + {\cal M}_{D^*\bar{D}} + \sum{\cal M}_{\rm BW} + {\cal M}_{\rm bk},
\end{align}
and the explicit forms, together with the Monte Carlo simulation procedure, will be presented below.

\subsubsection{Contribution from rescattering}

Following the preparatory analysis conducted in the previous sections, we
proceed to calculate the decay amplitude for the process $B^+ \to D^{*+} D^-
K^+$, incorporating the ${\cal M}_{\bar{D}^{*}K^{*}}$ and
${\cal M}_{D^*\bar{D}}$ rescattering  associated with the
$T^*_{\bar{c}\bar{s}0}(2870)^0$ and $\chi_{c1}(3872)$ states. For the
${\cal M}_{\bar{D}^{*}K^{*}}$, whose isospin quantum number $I$ has not been
experimentally confirmed, we independently compute the decay amplitudes
corresponding to the two possible isospin assignments, $I = 0$ and $I = 1$. The
contributions from the $D^{-}K^{+}$ channel are subsequently determined by
applying Eq.~(\ref{K^*D^*}) within the framework outlined in
Sec.~\ref{secHadronization}. In contrast, the analysis for ${\cal M}_{D^*\bar{D}}$ is
significantly simplified, as only the configuration with quantum numbers
$I^G(J^{PC}) = 0^+(1^{++})$ needs to be considered. Besides, to achieve better
agreement with the experimental data, the calculated amplitude is multiplied by
a phase factor. The corresponding mathematical expression can be formulated as
${\cal M}= e^{i\pi\alpha}{\cal M}^{J^P}$, where ${\cal M}^{J^P}$ denotes the
original computed amplitude in
Eq.~(\ref{amplitude}) and $\alpha$ is free parameter. Moreover, the cutoff parameter
$\Lambda$ is adjusted to achieve a satisfactory fit to the experimental data.
Due to the distinct interaction mechanisms involved in the $\bar{D}^{*}K^{*}$ and $D^*\bar{D}$ rescattering processes, it is appropriate to assign different
cutoff values for these two cases—specifically, 3.10~GeV for the former and
0.95~GeV for the latter. 

\subsubsection{Contribution from other resonance } \label{sec3b}

The experimental data analysis demonstrates that the invariant mass spectra of
$D^-K^+$ and $D^{*+}D^-$ include contributions from multiple resonances beyond
the $T^*_{\bar{c}\bar{s}0}(2870)^0$ and $\chi_{c1}(3872)$. In order to enhance
the consistency between our theoretical model and the experimentally observed
invariant mass distributions, we introduce Breit-Wigner resonances with spin
$J=1$ at approximately 2900~MeV, 4010~MeV, and 4300~MeV. These resonances are
incorporated to account for the observed signals corresponding to the
$T^*_{\bar{c}\bar{s}1}(2900)^0$, \\$\chi_{c1}(4010)$, and $h_{c}(4300)$ states
reported by the LHCb Collaboration. The resulting amplitude model can be
formulated as follows~\cite{LHCb:2024vfz}:
\begin{equation}
{\cal M}_{\rm BW}(m)=\frac{a_R\cdot e^{i\pi b_R}}{m_0^2-m^2-i m_0\Gamma(m)}.\label{Eq: BW}
\end{equation}
Here $a_{R}$ and $b_R$ are free parameters. The mass-dependent width is $\Gamma(m)=\Gamma_0\left(m_0 / m\right)\left(q / q_0\right)^{2 l+1} B_l^{\prime 2}\left(q, q_0, d\right)$, where $l$ corresponds to the angular momentum between the two decay products of the resonance $R$, $B_l^{\prime}\left(q, q_0, d\right)$ is the Blatt-Weisskopf barrier factor with $d=3.0$~GeV$^{-1}$, $q(q_0)$ denotes the momentum of the decay products in the rest frame of the resonance at the reconstructed mass $m$ (pole mass $m_0$) and $\Gamma_0$ is the width of the resonance. The predetermined masses($m_0$) and
widths($\Gamma_0$) are provided in Table~\ref{Tpara}.  
\renewcommand\tabcolsep{0.25cm}
\renewcommand{\arraystretch}{1.5}
\begin{table}[h!]
\centering
\caption{Masses and widths for the resonances involved. 
 Predetermined values are cited from Ref.~\cite{LHCb:2024vfz}.
\label{Tpara}}
\begin{tabular}{l ccc }
\hline
&Resonance & {Mass (MeV)} & Width (MeV) \\
\hline
Predetermiend 
&$T^*_{\bar{c}\bar{s}1}(2900)^0$ & 2914.0  & 128.0\\
&$\chi_{c1}(4010)$ & 4012.5  &62.7\\
&$h_{c}(4300)$ & 4307.3 &  58.0 \\\hline
\end{tabular}
\end{table}

\subsubsection{Contribution from background}
Parameterized background contributions are incorporated into the $D^-K^+$ and $D^{*+}D^-$ invariant mass distribution arising from the rescattering process discussed in Sec.~\ref{sec3a}. This background model is expressed as  
\begin{align}
{\cal M}_{\rm bk}(m)=c(m-M_{min})^a(M_{max}-m)^b,\label{Eq: background}
\end{align}
where the parameters for the background 12  and 23 are chosen as $(a, b, c)=(1.0,2.0,1.4)$
for $m_{D^-K^+}$ and $(a, b, c)=(1.0,6.0,3.0)$ for $m_{D^{*+}D^-}$, respectively, to fit the
experimental data.

\subsubsection{Monte Carlo simulations} \label{sec3c}

Based on the aforementioned preparations, Monte Carlo simulations are employed to calculate the invariant mass spectra of $D^-K^+$ and $D^{*+}D^-$. The resulting spectra are then compared with experimental data to investigate the structural properties of the $T^*_{\bar{c}\bar{s}0}(2870)^0$ and $\chi_{c1}(3872)$ states. 

The total decay width, incorporating all these contributions to the amplitude ${\cal M}$, can be expressed as
\begin{align}
{d\Gamma}&=\frac{(2\pi)^4}{2M_{B}}|{\cal M}|^2d\Phi_3,\label{Eq: IM} 
\end{align}
where $M_B$ denotes the mass of the initial $B^+$ meson. In this study, the three-body phase space $d\Phi_3$ in Eq.~(\ref{Eq: IM}) is generated using the GENEV code within the FAWL framework. This code utilizes the Monte Carlo method to simulate events corresponding to the three-body final state. The phase space is mathematically defined as
\begin{align}
R_3=(2 \pi)^5 d \Phi_3=\prod_i^3 \frac{d^3 p_i}{2 E_i} \delta^4\left(\sum_i^n p_i-P\right),
\end{align}
where $p_i$ and $E_i$ denote the momentum and energy of the final particle $i$, respectively, which are generated by the simulation code. By generating $5\times10^5$ Monte Carlo events, the event distribution is obtained, enabling the visualization of the invariant mass spectra for $m_{D^-K^+}$ and $m_{D^{*+}D^-}$.

\subsection{Invariant mass spectra with  rescatterings  $\bar{D}^{*}K^{*}$ and $D^*\bar{D}$} \label{sec3d}

\begin{figure}[htbp]
\centering
\includegraphics[width=1.03\linewidth, trim=0.45cm 0cm 0cm -0.3cm, clip]{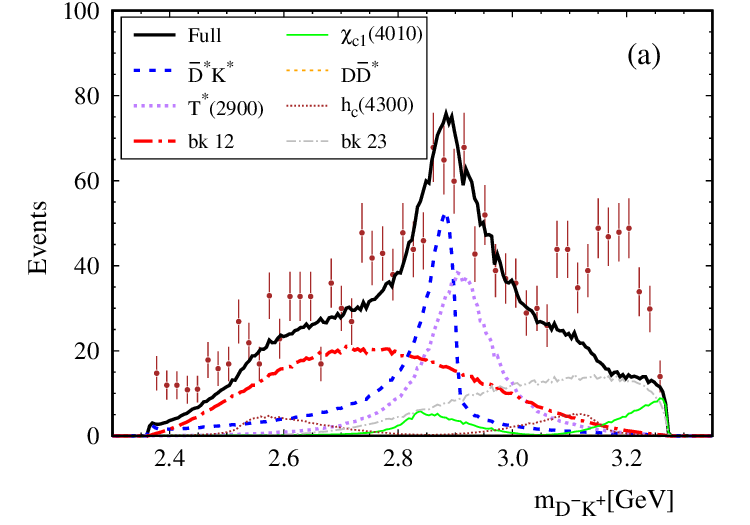}\\
\includegraphics[width=1.03\linewidth, trim=0.45cm 0cm 0cm 0.cm, clip]{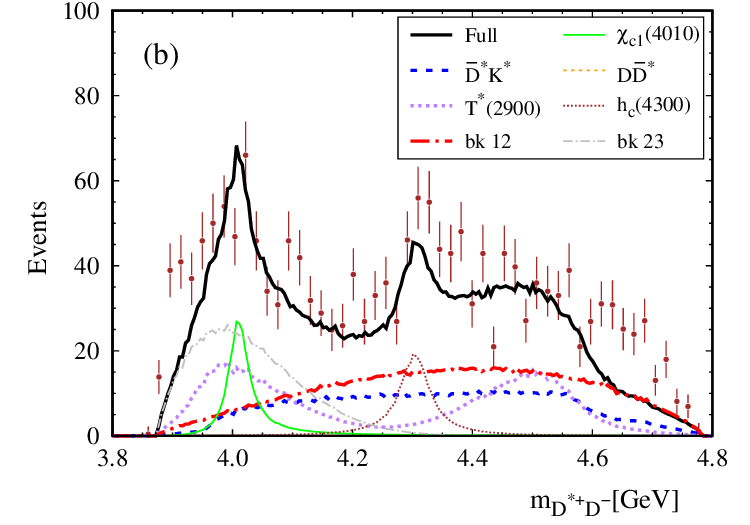}

\caption{Invariant mass spectra for the process $B^+ \to D^{*+} D^- K^+$ in the
(a) $D^-K^+$ and (b) $D^{*+}D^-$ channels. The thick solid black, thick dashed blue,
thick dotted purple, thick dash-dotted red, thin solid green, thin dashed
orange, thin dotted brown, and thin dash-dotted grey curves correspond to the
contributions from the full model, the $\bar{D}^{*}K^{*}$ molecular state (associated with
$T^*_{\bar{c}\bar{s}0}(2870)^0$), the $T^*_{\bar{c}\bar{s}1}(2900)^0$ resonance,
background in the $m_{D^-K^+}$ channel, the $\chi_{c1}(4010)$ resonance, the
$D^*\bar{D}$ molecular state (associated with $\chi_{c1}(3872)$), the $h_c(4300)$
resonance, and background in the $m_{D^{+}D^-}$ channel, respectively. Note that
the contribution from the $D^*\bar{D}$ molecular state is too small to be visible in
the figure.  The red points with error bars denote the LHCb experimental
data~\cite{LHCb:2024vfz}.  The theoretical curves are obtained using $5\times
10^5$ Monte Carlo simulations.}

  \label{Fig: overall}
\end{figure}

The invariant mass spectra including the above contributions are presented in
Fig.~\ref{Fig: overall}. Specifically, Figs.~\ref{Fig: overall}(a) and \ref{Fig:
overall}(b) show the invariant mass distributions for $m_{D^-K^+}$ and
$m_{D^{*+}D^-}$, respectively.  It is important to note that, due to the absence
of information on the total number of $B^+$ mesons, the absolute normalization
of the invariant mass spectra cannot be determined, as such data have not been
provided by the LHCb collaboration. To facilitate a meaningful comparison
between the theoretical predictions and the experimental results, the
theoretical decay distributions are rescaled to match the experimental spectra.
We emphasize that this scaling procedure allows us to extract only the relative
magnitudes of the coupling constants of $B^+$ decays $c_1$, $c_2$, and $c_3$, without
determining their absolute values. From the fit to the data, we find the relative
magnitude $c_2/c_1=6.67$ with  $c_3/c_1=\sqrt{2}$ which has been predetermined in hadronization.

As shown in Fig.~\ref{Fig: overall}(a), the LHCb experimental results reveal a
prominent resonant structure in the $D^-K^+$ invariant mass spectrum of the $B^+
\to D^{*+} D^- K^+$ decay process, centered around 2900~MeV. Previous theoretical
analysis indicates that this resonance lies close to the $\bar{D}^*K^*$ threshold
and can therefore be interpreted as a $\bar{D}^{*}K^{*}$ molecular state. Within
this framework, the calculated contribution of the $\bar{D}^{*}K^{*}$ molecular
state to the $D^-K^+$ invariant mass distribution is represented by the thick dashed blue line, which reproduces the experimentally observed
$T^*_{\bar{c}\bar{s}0}(2870)^0$ resonance. The corresponding $\bar{D}^{*}K^{*}$ molecular
state  with isopsin $I=0$ has a mass of  2886.75~MeV and  a width of 56.0~MeV. The strong agreement between our
theoretical results and the experimental data lends further support to the
interpretation of $T^*_{\bar{c}\bar{s}0}(2870)^0$ as a $\bar{D}^{*}K^{*}$ molecular
state.

Furthermore, as reported in Refs.~\cite{LHCb:2024vfz,LHCb:2020pxc}, the observed
enhancement near 2900~MeV is attributed to the combined contributions of two
states: $T^*_{\bar{c}\bar{s}0}(2870)^0$ and $T^*_{\bar{c}\bar{s}1}(2900)^0$. In our
calculation, the $\bar{D}^{*}K^{*}$ molecular state contributes predominantly
below 2900~MeV, whereas the experimental spectrum exhibits a broader peak. To
better reproduce the data, we include an additional Breit-Wigner resonance with
$J=1$ and a central mass near 2900~MeV, corresponding to the
$T^*_{\bar{c}\bar{s}1}(2900)^0$ state. Its parameterization follows Eq.~(\ref{Eq:
BW}), with the mass and width given in Table~\ref{Tpara}. The resulting
contribution is shown as the thick dotted purple line in Fig.~\ref{Fig: overall}.

With the inclusion of the $T^*_{\bar{c}\bar{s}1}(2900)^0$ component, the
theoretical prediction in the 2900~MeV region exhibits markedly improved
agreement with the LHCb data, as demonstrated by the thick black solid curve. This
outcome not only reinforces the interpretation of $T^*_{\bar{c}\bar{s}0}(2870)^0$
as an $S$-wave $\bar{D}^{*}K^{*}$ molecular state, but also confirms that the
$T^*_{\bar{c}\bar{s}1}(2900)^0$ contribution is essential for accurately
describing the resonance structure in this energy region.

Furthermore, two distinct peak-like structures are observed at approximately
4.01 and 4.30~GeV in Fig.~\ref{Fig: overall}(b), in agreement with the experimental data. These peaks are
modeled using Breit-Wigner resonances introduced to describe the
$\chi_{c1}(4010)$ and $h_c(4300)$ states, respectively. Experimental analysis of
the higher energy region reveals the presence of NR contributions~\cite{LHCb:2024vfz}. As
these NR components are not the primary focus of this study, no further fitting
has been performed on them. 

However, within the energy range above 3.0~GeV, our theoretical calculations do
not align well with the experimental data. As shown in  Fig.~\ref{Fig: overall}(a), the
experimental results exhibit an prominent enhancement structure in this
region, whereas no such peak is reproduced in our calculations. In our model,
the observed structure in this energy range should arise from contributions by
the $\chi_{c1}(3872)$, $\chi_{c1}(4010)$, and $h_c(4300)$ resonances, along with
background components from the $D^{*+}D^-$ invariant mass spectrum. However, the
contribution of $\chi_{c1}(3872)$ is found to be nearly negligible based on the
plotted results, while the other resonances generate only a mild slope-like
feature. And, although the contribution of $D^*\bar{D}$ interaction
 cannot be directly visualized, we still can obtain a pole from the
$D^*\bar{D}$ interaction, with a position at 3872.44 MeV. 

A similar phenomenon is also evident in Fig.~\ref{Fig: overall}(b).  From the
experimental results, a peak-like structure is observed at 3.87~GeV within the
energy range of 3.8 to 4.0~GeV in the $D^{*+}D^-$ invariant mass spectrum. 
According to our calculation results, the contribution from $D^*\bar{D}$
interaction is nearly negligible, which prevents the formation of the
enhancement at 3.87~GeV suggested by the experimental data. 

Such suppression of the $D^*\bar{D}$ interaction originates from the
hadronization mechanism discussed in Section~\ref{secHadronization}.  Since
there is only a single decay mechanism, as illustrated in Fig.~\ref{Fig:
Hadronization}, the rescattering processes involved $\bar{D}K$ and $D^*\bar{D}$ are
correlated, and thus the relative magnitudes of their contributions are fixed.
A reasonable description of the $\bar{D}^{*}K^{*}$ molecular state naturally leads
to a negligible $D^*\bar{D}$ contribution.  If the $D^*\bar{D}$ contribution were
artificially increased, the corresponding $\bar{D}^{*}K^{*}$ contribution would
become unphysically large.

\subsection{Invariant mass spectrum after introducing  Breit-Wigner resonance near 3872~MeV} \label{sec3e}

\begin{figure}[htbp]
\centering
\includegraphics[width=1.03\linewidth, trim=0.45cm 0cm 0cm 0.cm, clip]{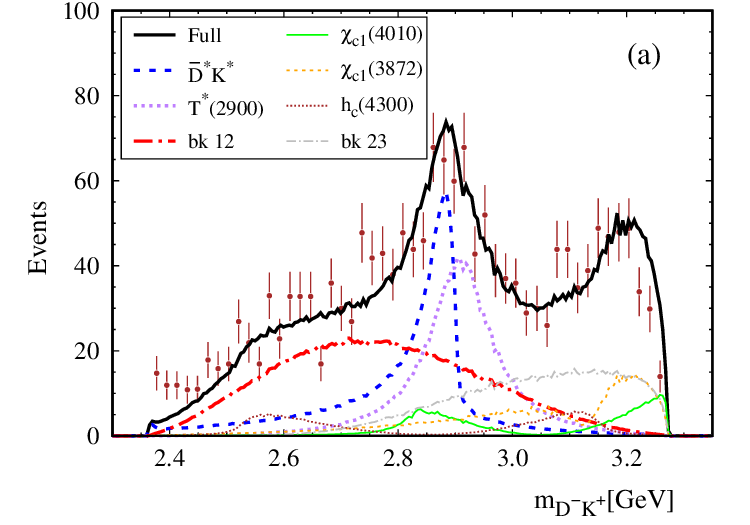}\\
\includegraphics[width=1.03\linewidth, trim=0.45cm 0cm 0cm 0.cm, clip]{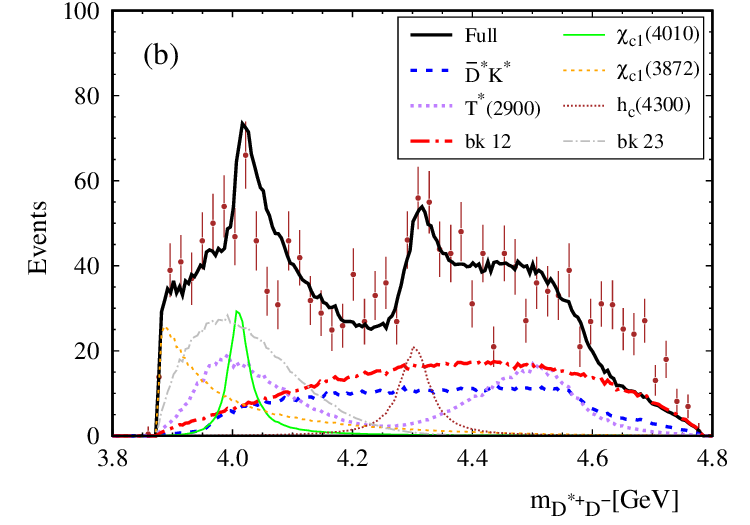}

\caption{Invariant mass spectra for the process $B^+ \to D^{*+} D^- K^+$ in the
(a) $D^-K^+$ and (b) $D^{*+}D^-$ channels after introducing a Breit–Wigner
$\chi_{c1}(3872)$ resonance near 3872~MeV. The contribution from the
$D\bar{D}^{*}$ molecular state is too small to be visible and is therefore not
shown in the figure; instead, the contribution from the Breit–Wigner
$\chi_{c1}(3872)$ resonance is presented. All other labels are the same as those
in Fig.~\ref{Fig: overall}.}

\label{Fig: overall2}
\end{figure}

In this subsection, we continue to treat $T^*_{\bar{c}\bar{s}0}(2870)^0$ as the
$D^{*-}K^{*+}$ molecular state, and a Breit-Wigner resonance with spin quantum
number $J=1$ is introduced to describe the structure near 3872~MeV, rather than
interpreting $\chi_{c1}(3872)$ as a $D\bar{D}^*$ molecular state. The resonance follows the
formulation given in Eq.~(\ref{Eq: BW}), with fixed mass $m_0 = 3871.64$~MeV and
width $\Gamma_0 = 1.19$~MeV~\cite{ParticleDataGroup:2024cfk}. After applying
Breit-Wigner fitting to $\chi_{c1}(3872)$, $\chi_{c1}(4010)$, $h_c(4300)$, and
$T^*_{\bar{c}\bar{s}1}(2900)^0$, we obtain the invariant mass spectra for
$D^-K^+$ and $D^{*+}D^-$ as in Fig.~\ref{Fig: overall2}.

Similar to Fig.~\ref{Fig: overall} in the previous subsection, the $D^-K^+$
invariant mass spectrum in Fig.~\ref{Fig: overall2}(a) exhibits a prominent peak
around 2.9~GeV. This structure arises primarily from the $\bar{D}^*K^*$ molecular
state, which can be interpreted as the $T^*_{\bar{c}\bar{s}0}(2870)^0$, together
with a contribution from the $T^*_{\bar{c}\bar{s}1}(2900)^0$ resonance. In the
$D^{*+}D^-$ invariant mass spectrum shown in Fig.~\ref{Fig: overall2}(b), two
additional peak-like structures are observed at approximately $4.01$~GeV and
$4.30$~GeV, which correspond to the $\chi_{c1}(4010)$ and $h_c(4300)$
resonances, respectively.

Compared with the spectra shown in Fig.~\ref{Fig: overall}, there are two newly
visible enhancements in Fig.~\ref{Fig: overall2}. In the $D^-K^+$ spectrum, an
additional peak appears around $3.2$~GeV, while in the $D^{*+}D^-$ spectrum,
another enhancement is observed near $3.87$~GeV. These newly observed structures
originate from the Breit–Wigner $\chi_{c1}(3872)$ resonance that is introduced near $3872$~MeV.
Their inclusion allows us to account for features in the experimental data that
are not explained solely by the previously considered molecular states and
higher-mass resonances.

LHCb's analysis shows that the measured branching ratios for $B^+ \to
\text{EFF}_{1{++}} K^+$ followed by $\text{EFF}_{1{++}} \to D^{* \pm} D^\mp$ are
higher than those for $B^+ \to \chi_{c1}(3872) K^+$ and $\chi_{c1}(3872) \to
D^{*0} \bar{D}^0$~\cite{LHCb:2024vfz,Belle:2008fma}. Due to the limited phase space
for $\chi_{c1}(3872) \to D^{* \pm} D^\mp$, this suggests a more complex decay
mechanism than a simple resonance. Furthermore, including $\chi_{c1}(3872)$ via
Eq.~(\ref{Eq: BW}) causes significant interference with other $1^{++}$ states,
making the model inconsistent~\cite{LHCb:2024vfz}. In this study, assuming
$\chi_{c1}(3872)$ as a $D^*\bar{D}$ molecular state necessitates constraints on
the production ratio of $D^{*+} D^-$ and $D^{*-} K^{*+}$ during the
hadronization process, thereby imposing a theoretical limitation on its relative
production yield. These findings may indicate that, within the framework of our
model, the structure observed near 3872~MeV in the $D^{*+} D^-$ invariant mass
spectrum is better attributed to an as-yet-undetermined resonance rather than to
$\chi_{c1}(3872)$.

\section{Summary}\label{sec4}

In this study, we carried out a comprehensive analysis of the invariant mass
spectra of $D^-K^+$ and $D^{*+}D^-$ in the decay process $B^+ \to D^{*+} D^- K^+$.
The analysis explicitly incorporates the rescattering mechanisms involving the
intermediate states $\chi_{c1}(3872)$ and $T^*_{\bar{c}\bar{s}0}(2870)^0$. Within
the molecular state framework, these intermediate states are interpreted as
$D^*\bar{D}$ and $\bar{D}^{*}K^{*}$ molecular states, respectively.

The hadronization mechanism imposes constraints on the possible quantum numbers
of these states. For the $\chi_{c1}(3872)$, if it is treated as a $D^*\bar{D}$
molecular state, its quantum numbers are restricted to $I^G(J^{PC}) =
0^+(1^{++})$, consistent with the assignments provided by the Particle Data
Group (PDG)~\cite{ParticleDataGroup:2024cfk}. In contrast, the
$T^*_{\bar{c}\bar{s}0}(2870)^0$, if interpreted as a $\bar{D}^{*}K^{*}$ molecular state,
is not subject to such quantum number constraints. Therefore, in our analysis,
we evaluated the contributions from both isospin components, $I=0$ and $I=1$, to
determine the overall contribution of the $T^*_{\bar{c}\bar{s}0}(2870)^0$ state.
This approach also naturally establishes the relative relationship between the
contributions of the two molecular states.

Subsequently, using theoretical calculations based on the quasipotential
Bethe-Salpeter equation approach, we computed the decay amplitudes and invariant
mass distributions, which were then compared with experimental data from the
LHCb collaboration. 
To achieve better agreement between theoretical predictions and experimental
observations, we incorporated the experimentally observed intermediate states
$T^*_{\bar{c}\bar{s}1}(2900)^0$, $\chi_{c1}(4010)$, and $h_c(4300)$ into our model
as Breit-Wigner resonances with spin quantum number $J=1$.

We first examined the $D^-K^+$ invariant mass spectrum in the decay process $B^+
\to D^{*+} D^- K^+$. The interaction of $\bar{D}^{*}K^{*}$ generates a pronounced peak at
2886.75~MeV with a width of 56.0~MeV, which can be associated with the
$T^*_{\bar{c}\bar{s}0}(2870)^0$ state reported by the LHCb collaboration. Upon
including the $T^*_{\bar{c}\bar{s}1}(2900)^0$ resonance, the theoretical
prediction around 2.9~GeV shows significantly improved agreement with the
experimental data. However, noticeable discrepancies remain in the higher energy
region, particularly in the 3.0$-$3.2~GeV range. While the experimental spectrum
exhibits a distinct peak-like structure near 3.2~GeV, our model fails to
reproduce this feature. This discrepancy mainly originates from the negligible
contribution of the $\chi_{c1}(3872)$ state within the current theoretical
framework.

A similar situation occurs in the $D^{*+} D^-$ invariant mass spectrum.
Theoretical calculations do not accurately reproduce the peak-like structure
around 3.87~GeV, which again can be attributed to the minimal contribution of
$\chi_{c1}(3872)$. 
Outside of this energy region, the agreement between the
model and experimental data is relatively satisfactory. The observed peak-like
structures near 4.01 and 4.30~GeV in the $D^{*+} D^-$ spectrum can be
attributed to the introduced Breit-Wigner resonances $\chi_{c1}(4010)$ and
$h_c(4300)$, respectively, highlighting the importance of these states in
accurately describing the higher-mass region of the decay spectra.

In our calculation, the assumption of a $D^*\bar{D}$ molecular configuration for
$\chi_{c1}(3872)$ fails to reproduce the experimental features observed near
3.87~GeV in the $D^{+}D^-$ invariant mass spectrum. In the experimental analysis
of this decay process~\cite{LHCb:2024vfz}, however, the LHCb collaboration
reported that the enhancement in this energy region can be effectively described
by a $\chi_{c1}(3872)$ model within the Breit-Wigner formalism. To further
clarify this issue, we adopted the data-fitting methodology employed by LHCb and
introduced a $J=1$ Breit-Wigner parameterization to represent the
$\chi_{c1}(3872)$ resonance near 3.87GeV. The results demonstrate that, after
incorporating this Breit-Wigner resonance, the theoretical simulation achieves a
significantly improved agreement with experimental observations. In particular,
both the peak-like structure near 3.2GeV in the $D^-K^+$ invariant mass spectrum
and the resonance-like feature around 3.87~GeV in the $D^{*+}D^-$ spectrum are
more accurately reproduced. These findings suggest that, within the molecular
state framework, the structure observed near 3.87~GeV may not be fully explained
as the $\chi_{c1}(3872)$. 

\vskip 10pt

\noindent {\bf Data Availability Statement} This manuscript has no associated
data or the data will not be deposited. [Authors' comment: This is a theoretical
study and no external data are associated with this work.]

%


\begin{thebibliography}{23}%
\bibitem{LHCb:2024vfz}
R.~Aaij \textit{et al.} [LHCb],
``Observation of New Charmonium or Charmoniumlike States in $B^+\to D^{*\pm}D^\mp K^+$ Decays,''
Phys. Rev. Lett. \textbf{133} (2024) no.13, 131902



\bibitem{LHCb:2020bls}
R.~Aaij \textit{et al.} [LHCb],
``A model-independent study of resonant structure in $B^+\to D^+D^-K^+$ decays,''
Phys. Rev. Lett. \textbf{125} (2020), 242001

\bibitem{LHCb:2020pxc}
R.~Aaij \textit{et al.} [LHCb],
``Amplitude analysis of the $B^+\to D^+D^-K^+$ decay,''
Phys. Rev. D \textbf{102} (2020), 112003

\bibitem{Liu:2020nil}
M.~Z.~Liu, J.~J.~Xie and L.~S.~Geng,
``$X_0(2866)$ as a $D^*\bar{K}^*$ molecular state,''
Phys. Rev. D \textbf{102} (2020) no.9, 091502

\bibitem{Ke:2022ocs}
H.~W.~Ke, Y.~F.~Shi, X.~H.~Liu and X.~Q.~Li,
``Possible molecular states of D{\textasciimacron}*K* (D*K*) and new exotic states X0(2900), X1(2900), Tcs0a(2900)0 and Tcs0a(2900)++,''
Phys. Rev. D \textbf{106} (2022) no.11, 114032

\bibitem{Yu:2023avh}
Z.~Yu, Q.~Wu and D.~Y.~Chen,
``$X_0(2900)$ and its spin partners productions in the $B^+$ decay processes,''
Eur. Phys. J. C \textbf{84} (2024) no.9, 985

\bibitem{Ikeno:2022jbb}
N.~Ikeno, M.~Bayar and E.~Oset,
``Molecular states of D*D*K{\textasciimacron}* nature,''
Phys. Rev. D \textbf{107} (2023) no.3, 034006



\bibitem{Hu:2020mxp}
M.~W.~Hu, X.~Y.~Lao, P.~Ling and Q.~Wang,
``$X_0$(2900) and its heavy quark spin partners in molecular picture,''
Chin. Phys. C \textbf{45} (2021) no.2, 021003

\bibitem{Yang:2024coj}
Z.~Y.~Yang, Q.~Wang and W.~Chen,
``Production and decay of the X0(2900) state with different interpretations,''
Phys. Rev. D \textbf{111} (2025) no.7, 076030

\bibitem{Chen:2020aos}
H.~X.~Chen, W.~Chen, R.~R.~Dong and N.~Su,
``$X_0$(2900) and $X_1$(2900): Hadronic Molecules or Compact Tetraquarks,''
Chin. Phys. Lett. \textbf{37} (2020) no.10, 101201



\bibitem{Liu:2022hbk}
F.~X.~Liu, R.~H.~Ni, X.~H.~Zhong and Q.~Zhao,
``Charmed-strange tetraquarks and their decays in the potential quark model,''
Phys. Rev. D \textbf{107} (2023) no.9, 096020

\bibitem{Wang:2020xyc}
Z.~G.~Wang,
``Analysis of the $X_0(2900)$ as the scalar tetraquark state via the QCD sum rules,''
Int. J. Mod. Phys. A \textbf{35} (2020) no.30, 2050187

\bibitem{Zhang:2020oze}
J.~R.~Zhang,
``Open-charm tetraquark candidate: Note on $X_0$(2900),''
Phys. Rev. D \textbf{103} (2021) no.5, 054019

\bibitem{Lian:2023cgs}
D.~K.~Lian, W.~Chen, H.~X.~Chen, L.~Y.~Dai and T.~G.~Steele,
``Strong decays of $T^a_{c{\bar{s}0}}(2900)^{++/0}$ as a fully open-flavor tetraquark state,''
Eur. Phys. J. C \textbf{84} (2024) no.1, 1-10


\bibitem{Albuquerque:2021svg}
R.~M.~Albuquerque, S.~Narison, D.~Rabetiarivony and G.~Randriamanatrika,
``The New Charm-Strange Resonances in the $D^- K^+$ Channel,''
Nucl. Part. Phys. Proc. \textbf{312-317} (2021), 125-129




\bibitem{Molina:2023ghu}
R.~Molina and E.~Oset,
``The Tcs (2900) as a threshold effect from the interaction of the D*K*, D*s{\ensuremath{\rho}} channels,''
EPJ Web Conf. \textbf{291} (2024), 03010

\bibitem{Burns:2020epm}
T.~J.~Burns and E.~S.~Swanson,
``Kinematical cusp and resonance interpretations of the $X(2900)$,''
Phys. Lett. B \textbf{813} (2021), 136057

\bibitem{Molina:2010tx}
R.~Molina, T.~Branz and E.~Oset,
``A new interpretation for the $D^*_{s2}(2573)$ and the prediction of novel exotic charmed mesons,''
Phys. Rev. D \textbf{82} (2010), 014010

\bibitem{Chen:2021erj}
H.~X.~Chen,
``Hadronic molecules in B decays,''
Phys. Rev. D \textbf{105} (2022) no.9, 094003


\bibitem{Wang:2021lwy}
B.~Wang and S.~L.~Zhu,
``How to understand the X(2900)?,''
Eur. Phys. J. C \textbf{82} (2022) no.5, 419

\bibitem{He:2020btl}
J.~He and D.~Y.~Chen,
``Molecular picture for $X_0(2900)$ and $X_1(2900)$,''
Chin. Phys. C \textbf{45} (2021) no.6, 063102

\bibitem{Kong:2021ohg}
S.~Y.~Kong, J.~T.~Zhu, D.~Song and J.~He,
``Heavy-strange meson molecules and possible candidates Ds0*(2317), Ds1(2460), and X0(2900),''
Phys. Rev. D \textbf{104} (2021) no.9, 094012

\bibitem{Ding:2024dif}
Z.~M.~Ding, Q.~Huang and J.~He,
``$X_0(2900)$ and $\chi _{c0}(3930)$ in process $B^+\rightarrow D^+ D^- K^+$,''
Eur. Phys. J. C \textbf{84} (2024) no.8, 822

\bibitem{Liu:2009qhy}
X.~Liu, Z.~G.~Luo, Y.~R.~Liu and S.~L.~Zhu,
``X(3872) and Other Possible Heavy Molecular States,''
Eur. Phys. J. C \textbf{61} (2009), 411-428

\bibitem{AlFiky:2005jd}
M.~T.~AlFiky, F.~Gabbiani and A.~A.~Petrov,
``X(3872): Hadronic molecules in effective field theory,''
Phys. Lett. B \textbf{640} (2006), 238-245

\bibitem{Wang:2013daa}
Z.~G.~Wang and T.~Huang,
``Possible assignments of the $X(3872)$, $Z_c(3900)$ and $Z_b(10610)$ as axial-vector molecular states,''
Eur. Phys. J. C \textbf{74} (2014) no.5, 2891

\bibitem{Dong:2009yp}
Y.~Dong, A.~Faessler, T.~Gutsche, S.~Kovalenko and V.~E.~Lyubovitskij,
``X(3872) as a hadronic molecule and its decays to charmonium states and pions,''
Phys. Rev. D \textbf{79} (2009), 094013

\bibitem{He:2014nya}
J.~He,
``Study of the $B\bar{B}^*/D\bar{D}^*$ bound states in a Bethe-Salpeter approach,''
Phys. Rev. D \textbf{90} (2014) no.7, 076008

\bibitem{Ding:2020dio}
Z.~M.~Ding, H.~Y.~Jiang and J.~He,
``Molecular states from $D^{(*)}\bar{D}^{(*)}/B^{(*)}\bar{B}^{(*)}$ and $D^{(*)}D^{(*)}/\bar{B}^{(*)}\bar{B}^{(*)}$ interactions,''
Eur. Phys. J. C \textbf{80} (2020) no.12, 1179


\bibitem{Oset:2016lyh}
E.~Oset, W.~H.~Liang, M.~Bayar, J.~J.~Xie, L.~R.~Dai, M.~Albaladejo, M.~Nielsen, T.~Sekihara, F.~Navarra and L.~Roca, \textit{et al.}
``Weak decays of heavy hadrons into dynamically generated resonances,''
Int. J. Mod. Phys. E \textbf{25} (2016), 1630001

\bibitem{Braaten:2004fk}
E.~Braaten, M.~Kusunoki and S.~Nussinov,
``Production of the X(3870) in B meson decay by the coalescence of charm mesons,''
Phys. Rev. Lett. \textbf{93} (2004), 162001

\bibitem{Huang:2020ptc}
Y.~Huang, J.~X.~Lu, J.~J.~Xie and L.~S.~Geng,
``Strong decays of ${\bar{D}}^{*}K^{*}$ molecules and the newly observed $X_{0,1}$ states,''
Eur. Phys. J. C \textbf{80} (2020) no.10, 973


\bibitem{Dai:2022htx}
L.~R.~Dai, R.~Molina and E.~Oset,
``Looking for the exotic X0(2866) and its JP=1+ partner in the B{\textasciimacron}0{\textrightarrow}D(*)+K-K(*)0 reactions,''
Phys. Rev. D \textbf{105} (2022) no.9, 096022

\bibitem{Bando:1984ej}
M.~Bando, T.~Kugo, S.~Uehara, K.~Yamawaki and T.~Yanagida,
``Is rho Meson a Dynamical Gauge Boson of Hidden Local Symmetry?,''
Phys. Rev. Lett. \textbf{54} (1985), 1215

\bibitem{Bando:1987br}
M.~Bando, T.~Kugo and K.~Yamawaki,
``Nonlinear Realization and Hidden Local Symmetries,''
Phys. Rept. \textbf{164} (1988), 217-314

\bibitem{Nagahiro:2008cv}
H.~Nagahiro, L.~Roca, A.~Hosaka and E.~Oset,
``Hidden gauge formalism for the radiative decays of axial-vector mesons,''
Phys. Rev. D \textbf{79} (2009), 014015
[arXiv:0809.0943 [hep-ph]].

\bibitem{Cheng:1992xi}
H.~Y.~Cheng, C.~Y.~Cheung, G.~L.~Lin, Y.~C.~Lin, T.~M.~Yan and H.~L.~Yu,
``Chiral Lagrangians for radiative decays of heavy hadrons,''
Phys. Rev. D \textbf{47} (1993), 1030-1042

\bibitem{Yan:1992gz}
T.~M.~Yan, H.~Y.~Cheng, C.~Y.~Cheung, G.~L.~Lin, Y.~C.~Lin and H.~L.~Yu,
``Heavy quark symmetry and chiral dynamics,''
Phys. Rev. D \textbf{46} (1992), 1148-1164

\bibitem{Wise:1992hn}
M.~B.~Wise,
``Chiral perturbation theory for hadrons containing a heavy quark,''
Phys. Rev. D \textbf{45} (1992) no.7, R2188

\bibitem{Burdman:1992gh}
G.~Burdman and J.~F.~Donoghue,
``Union of chiral and heavy quark symmetries,''
Phys. Lett. B \textbf{280} (1992), 287-291

\bibitem{Casalbuoni:1996pg}
R.~Casalbuoni, A.~Deandrea, N.~Di Bartolomeo, R.~Gatto, F.~Feruglio and G.~Nardulli,
``Phenomenology of heavy meson chiral Lagrangians,''
Phys. Rept. \textbf{281} (1997), 145-238

\bibitem{Falk:1992cx}
A.~F.~Falk and M.~E.~Luke,
``Strong decays of excited heavy mesons in chiral perturbation theory,''
Phys. Lett. B \textbf{292} (1992), 119-127

\bibitem{Isola:2003fh}
C.~Isola, M.~Ladisa, G.~Nardulli and P.~Santorelli,
``Charming penguins in B ---{\ensuremath{>}} K* pi, K(rho, omega, phi) decays,''
Phys. Rev. D \textbf{68} (2003), 114001

\bibitem{Chen:2019asm}
R.~Chen, Z.~F.~Sun, X.~Liu and S.~L.~Zhu,
``Strong LHCb evidence supporting the existence of the hidden-charm molecular pentaquarks,''
Phys. Rev. D \textbf{100} (2019) no.1, 011502

\bibitem{Ding:2023yuo}
Z.~m.~Ding and J.~He,
``Combined analysis on nature of $X(3960)$, $\chi_{c0}(3930)$, and $X_0(4140)$,''
Eur. Phys. J. C \textbf{83} (2023) no.9, 806

\bibitem{Oh:2000qr}
Y.~s.~Oh, T.~Song and S.~H.~Lee,
``J / psi absorption by pi and rho mesons in meson exchange model with anomalous parity interactions,''
Phys. Rev. C \textbf{63} (2001), 034901

\bibitem{He:2015mja}
J.~He,
``The $Z_c(3900)$ as a resonance from the $D\bar{D}^*$ interaction,''
Phys. Rev. D \textbf{92} (2015) no.3, 034004

\bibitem{He:2017lhy}
J.~He and D.~Y.~Chen,
``$Z_c(3900)/Z_c(3885)$ as a virtual state from $\pi J/\psi-\bar{D}^*D$ interaction,''
Eur. Phys. J. C \textbf{78} (2018) no.2, 94

\bibitem{He:2015yva}
J.~He,
``Internal structures of the nucleon resonances N(1875) and N(2120),''
Phys. Rev. C \textbf{91} (2015) no.1, 018201

\bibitem{He:2017aps}
J.~He,
``Nucleon resonances $N(1875)$ and $N(2100)$ as strange partners of LHCb pentaquarks,''
Phys. Rev. D \textbf{95} (2017) no.7, 074031

\bibitem{He:2015cea}
J.~He,
``$\bar{D}\Sigma^*_c$ and $\bar{D}^*\Sigma_c$ interactions and the LHCb hidden-charmed pentaquarks,''
Phys. Lett. B \textbf{753} (2016), 547-551


\bibitem{He:2019ify}
J.~He,
``Study of $P_c(4457)$, $P_c(4440)$, and $P_c(4312)$ in a quasipotential Bethe-Salpeter equation approach,''
Eur. Phys. J. C \textbf{79} (2019) no.5, 393

\bibitem{Gross:2008ps}
F.~Gross and A.~Stadler,
``Covariant spectator theory of np scattering: Phase shifts obtained from precision fits to data below 350-MeV,''
Phys. Rev. C \textbf{78} (2008), 014005

\bibitem{ParticleDataGroup:2024cfk}
S.~Navas \textit{et al.} [Particle Data Group],
``Review of particle physics,''
Phys. Rev. D \textbf{110} (2024) no.3, 030001

\bibitem{Belle:2008fma}
T.~Aushev \textit{et al.} [Belle],
``Study of the $B\to X(3872)(\to D^{*0}\bar{D}^0) K$ decay,''
Phys. Rev. D \textbf{81} (2010), 031103



\end{thebibliography}

\end{document}